 \documentclass[longauth]{aa}  

\usepackage{hyperref}

\usepackage{amsmath}
\usepackage{graphicx}
\usepackage{graphics}

\usepackage{natbib}{}
\usepackage{xcolor}
\usepackage{url}

\bibpunct{(}{)}{;}{a}{}{,}

\def\teff{\ifmmode T_{\rm eff} \else $T_{\mathrm{eff}}$\fi}

\def\ltsima{$\buildrel<\over\sim$}
\def\lsim{\lower.5ex\hbox{\ltsima}}

\newcommand{\hii}{H~{\sc ii}}
\newcommand{\ha}{\ifmmode {\rm H}\alpha \else H$\alpha$\fi}
\newcommand{\hb}{\ifmmode {\rm H}\beta \else H$\beta$\fi}
\newcommand{\hg}{\ifmmode {\rm H}\gamma \else H$\gamma$\fi}
\newcommand{\lya}{\ifmmode {\rm Ly}\alpha \else Ly$\alpha$\fi}

\newcommand{\ebv}{\ifmmode E_{\rm B-V} \else $E_{\rm B-V}$\fi}
\newcommand{\av}{\ifmmode A_{\rm V} \else $A_{\rm V}$\fi}

\def\cmc{cm$^{-3}$}

\def\msun{\ifmmode M_{\odot} \else M$_{\odot}$\fi}
\def\msunyr{\ifmmode M_{\odot} {\rm yr}^{-1} \else M$_{\odot}$ yr$^{-1}$\fi}
\def\zsun{\ifmmode Z_{\odot} \else Z$_{\odot}$\fi}

\def\lsun{\ifmmode L_{\odot} \else L$_{\odot}$\fi}

\def\mup{\ifmmode M_{\rm up} \else M$_{\rm up}$\fi}
\def\mlow{\ifmmode M_{\rm low} \else M$_{\rm low}$\fi}
\def\aap{A\&A}

\def\apj{ApJ}
\def\apjl{ApJL}
\def\apjs{ApJS}
\def\mnras{MNRAS}
\def\pasp{PASP}
\newcommand{\oh}{\ifmmode 12 + \log({\rm O/H}) \else$12 + \log({\rm
O/H})$\fi}
\newcommand{\nii}{[N~{\sc ii}]}

\newcommand{\oii}{[O~{\sc ii}]}
\newcommand{\oiii}{[O~{\sc iii}]}

\def\Nii{[N~{\sc ii}] $\lambda$6584}
\def\Sii{[S~{\sc ii}] $\lambda\lambda$6717,6731}

\def\Oii{[O~{\sc ii}] $\lambda\lambda$3727,3729}
\def\Oiii{[O~{\sc iii}] $\lambda\lambda$4959,5007}

\def\Oiiit{[O~{\sc iii}] $\lambda 4363$}
\newcommand{\Neiii}{[Ne~{\sc iii}] $\lambda$3869}

\def\flyf{\ifmmode f_{\rm Lyf} \else $f_{\rm Lyf}$\fi}
\def\pz{\ifmmode P(z) \else $P(z)$\fi}
\def\ki2{\ifmmode \chi^2 \else $\chi^2$\fi}
\def\zphot{\ifmmode z_{\rm phot} \else $z_{\rm phot}$\fi}

\newcommand{\xphot}{\ifmmode x_\gamma \else $v_\gamma$\fi}
\newcommand{\xobs}{\ifmmode x_{\rm obs} \else $x_{\rm obs}$\fi}
\newcommand{\xcmf}{\ifmmode x_{\rm CMF} \else $x_{\rm CMF}$\fi}
\newcommand{\vexp}{\ifmmode V_{\rm exp} \else $V_{\rm exp}$\fi}
\newcommand{\vmax}{\ifmmode V_{\rm max} \else $V_{\rm max}$\fi}
\newcommand{\nh}{\ifmmode N_{\rm HI} \else $N_{\rm HI}$\fi}
\newcommand{\dv}{\ifmmode \Delta v({\rm em-abs}) \else $\Delta v({\rm em}-{\rm abs})$\fi}

\def\fesc{\ifmmode f_{\rm esc} \else $f_{\rm esc}$\fi}
\def\fescrel{\ifmmode f_{\rm esc,rel} \else $f_{\rm esc,rel}$\fi}

\def\frellya{\ifmmode f^{\rm rel}_{\rm{Ly}\alpha} \else $f^{\rm rel}_{\rm{Ly}\alpha}$\fi}

\def\hii{H{\sc ii}}

\newcommand{\mstar}{\ifmmode M_\star \else $M_\star$\fi}
\newcommand{\muv}{\ifmmode M_{1500} \else $M_{1500}$\fi}
\newcommand{\auv}{\ifmmode A_{\rm UV} \else $A_{\rm UV}$\fi}
\newcommand{\luv}{\ifmmode L_{\rm UV} \else $L_{\rm UV}$\fi}
\newcommand{\lir}{\ifmmode L_{\rm IR} \else $L_{\rm IR}$\fi}
\newcommand{\lbol}{\ifmmode L_{\rm bol} \else $L_{\rm bol}$\fi}
\newcommand{\liruv}{\ifmmode L_{\rm IR+UV} \else $L_{\rm IR+UV}$\fi}
\newcommand{\liroveruv}{\ifmmode L_{\rm IR}/L_{\rm UV} \else $L_{\rm IR}/L_{\rm UV}$\fi}
\newcommand{\nlyc}{\ifmmode N_{\rm Lyc} \else $N_{\rm Lyc} $\fi}
\newcommand{\rholyc}{\ifmmode \rho_{\rm Lyc} \else $\rho_{\rm Lyc} $\fi}
\newcommand{\chion}{\ifmmode \xi_{\rm ion} \else $\xi_{\rm ion}$\fi}
\newcommand{\chioncorr}{\ifmmode \xi_{\rm ion}^0 \else $\xi_{\rm ion}^0$\fi}

\newcommand{\Niiiuv}{N~{\sc iii}] $\lambda$1750}
\newcommand{\Nivuv}{N~{\sc iv}] $\lambda$1486}

\begin{document}

\title{Nitrogen abundances in star-forming galaxies 2.2 Gyr after the Big Bang are not elevated}
\subtitle{}
\author{D. Schaerer\inst{1,2}, 
Y.I. Izotov\inst{3},
R. Marques-Chaves\inst{1}, 
C. C. Steidel\inst{4},
N. Reddy \inst{5},
A. E. Shapley\inst{6},
S. Mascia\inst{7},
J. Chisholm\inst{8},
S. R. Flury\inst{9},
N. Guseva\inst{3},
T. Heckman\inst{10,11},
A. Henry\inst{12},
A.K. Inoue\inst{13,14},
I. Jung\inst{12},
H. Kusakabe\inst{15},
K. Mawatari\inst{13,14},
P. Oesch\inst{1,16},
G. \"Ostlin\inst{17},
L. Pentericci\inst{18},
N. Roy\inst{10,11},
A. Saldana-Lopez\inst{17},
R. Sato\inst{14},
E. Vanzella\inst{19},
A. Verhamme\inst{1},
B. Wang\inst{20,21,22}}
  \institute{Department of Astronomy, University of Geneva, Chemin Pegasi 51, 1290 Versoix, Switzerland
\and CNRS, IRAP, 14 Avenue E. Belin, 31400 Toulouse, France
\and Bogolyubov Institute for Theoretical Physics,
National Academy of Sciences of Ukraine, 14-b Metrolohichna str., Kyiv,
03143, Ukraine
\and Cahill Center for Astronomy and Astrophysics, California Institute of Technology, 1216 East California Boulevard., MS 249-17, Pasadena, CA 91125, USA
\and Department of Physics and Astronomy, University of California, Riverside, 900 University Avenue, Riverside, CA, 92521, USA
\and Department of Physics and Astronomy, University of California, Los Angeles, 430 Portola Plaza, Los Angeles, CA 90095, USA
\and Institute of Science and Technology Austria (ISTA), Am Campus 1, 3400 Klosterneuburg, Austria
\and Department of Astronomy, The University of Texas at Austin, Austin, TX 78712, USA
\and Institute for Astronomy, University of Edinburgh, Royal Observatory, Edinburgh, EH9 3HJ, UK
\and Center for Astrophysical Sciences, Department of Physics \& Astronomy, Johns Hopkins University, Baltimore, MD 21218, USA 
\and School of Earth and Space Exploration, Arizona State University, Tempe, AZ 85287, USA
\and  Space Telescope Science Institute, 3700 San Martin Dr., Baltimore, MD 21218, USA
\and Waseda Research Institute for Science and Engineering, Faculty of Science and Engineering, Waseda University, 3-4-1 Okubo, Shinjuku, Tokyo 169-8555, Japan; 
\and Department of Physics, School of Advanced Science and Engineering, Faculty of Science and Engineering, Waseda University, 3-4-1 Okubo, Shinjuku, Tokyo 169-8555, Japan
\and Graduate School of Arts and Sciences, The University of Tokyo, 3-8-1 Komaba, Meguro-ku, Tokyo 153-8902, Japan
\and Cosmic Dawn Center (DAWN), Niels Bohr Institute, University of Copenhagen, Jagtvej 128, K{\o}benhavn N, DK-2200, Denmark
\and Department of Astronomy, The Oskar Klein Centre, Stockholm University, AlbaNova, SE-10691 Stockholm, Sweden
\and INAF - Osservatorio Astronomico di Roma, via Frascati 33, 00078, Monteporzio Catone, Italy
\and INAF - Osservatorio di Astrofisica e Scienza dello Spazio di Bologna, Via Gobetti 93/3, 40129, Bologna, Italy
\and Department of Astronomy \& Astrophysics, The Pennsylvania State University, University Park, PA 16802, USA
\and Institute for Computational \& Data Sciences, The Pennsylvania State University, University Park, PA 16802, USA
\and Institute for Gravitation and the Cosmos, The Pennsylvania State University, University Park, PA 16802, USA
}

\authorrunning{Schaerer et al.}
\titlerunning{Normal N/O abundances at redshift $z \sim 3$}

\date{Received date; accepted date}


\abstract{
Using deep medium-resolution JWST rest-optical spectra of a sample of  typical star-forming galaxies (Lyman break galaxies and Lyman-$\alpha$ emitters) from the LyC22 survey at $z \sim 3$, we determined the nebular abundances of N, O, and Ne relative to H for a subsample of 25 objects with the direct method, based on auroral \Oiiit\ line detections. Our measurements  increases the number of accurate N/O determinations at $z \sim 2-4$ using  a homogeneous approach. We found a mean value of $\log({\rm N/O})=-1.29^{+0.25}_{-0.21} $ over a metallicity range $\oh=7.56$ to 8.44. The observed N/O ratio and scatter are indistinguishable from that observed in low-$z$ galaxies and \hii\ regions over the same metallicity range, showing thus no redshift evolution of N/O for typical galaxies over a significant fraction of cosmic time.
We also show that typical $z \sim 3$ galaxies show a similar offset in the BPT diagram as galaxies from the low-$z$ Lyman Continuum Survey (LzLCS), when compared to the average of SDSS galaxies, and show that this offset is not due to enhanced nitrogen abundances. Our results establish a basis for future studies of the evolution of N and O at higher redshifts.
}


 \maketitle

\section{Introduction}
\label{s_intro}

The cosmic origin of nitrogen, an element that is abundant in the cosmos and important for life, has been studied for a long time \citep[e.g.][]{Edmunds1978Nitrogen-synthe,Henry2000On-the-Cosmic-O}
and its observed evolution in Milky Way stars, \hii\ regions, and low-redshift galaxies is fairly well understood and reproduced by chemical evolution models \citep[e.g.][]{Vincenzo2018Evolution-of-N/}. Recent JWST observations have revealed a new category of rare objects showing emission lines of \Nivuv\ and \Niiiuv\ in the rest-UV (hereafter named N-emitters), which are found to show highly supersolar N/O abundance ratios (by factors $\ga 2$ in several cases) yet  subsolar O/H \citep[see compilation of][]{Ji2025Connecting-JWST}, although not exclusively, as the discovery of new N-emitters shows \citep{Morel2025Discovery-of-ne}.

The nature of these objects, including the $z=10.6$ galaxy GN-z11, the source of enrichment, or other mechanisms leading to a high N/O ratio, are currently a very active topic of debate.

To place these objects into context, a major question remains, however: What is the nitrogen abundance and N/O ratio of typical galaxies at high redshift ?  From stacks of $\sim 1000$ low-resolution (PRISM) spectra taken with NIRSpec/JWST, \cite{Hayes2025On-the-Average-} suggest a possible increase of N/O at $z>4$ from the UV lines, compared to the abundances observed at low redshift. However, the determination of chemical abundances from these spectra is very uncertain.
Other studies have used deep medium-resolution rest-optical JWST spectra to measure auroral lines for accurate determinations of metallicities (O/H) and the \Nii\ line to then determine the N-abundance.  So far this has been achieved for a small number of galaxies ($\sim 30$) spanning a large redshift range ($z \sim 1.8-6.3$), finding a spread of N/O ratios and only few objects ($\sim 6$) with supersolar N/O \citep[][]{Arellano-Cordova2024The-JWST-EXCELS,Scholte2025The-JWST-EXCELS,Stiavelli2025What-Can-We-Lea,Marques-Chaves2025Extremely-UV-br}, including lensed galaxies like RXCJ2248 and the Sunburst arc  \citep{Pascale2023Nitrogen-enrich,Topping2024Metal-poor-star,Welch2025The-Sunburst-Ar,Berg2025A-Fleeting-GLIM}.
Another recent study has analysed rest-optical spectra from the MARTA program and other JWST archival data, suggesting that the median N/O abundance ratio of $z>1$ galaxies could be elevated by $\sim 0.18$ dex at fixed metallicity (O/H) compared to the average N/O--O/H relations established locally \citep{Cataldi2025Tracing-Nitroge}.

To address these questions and establish the typical N-abundance in ``typical'' star-forming galaxies (SFGs) at high redshift, we here present an analysis of 117 deep JWST spectra of Lyman break galaxies (LBGs) and Lyman-$\alpha$ emitters (LAEs) at $z \sim 3$ drawn from the LyC22 survey (GO 1869, PI Schaerer), from which we determine accurate chemical abundances of N, O, and Ne relative to H using the direct method for 25 galaxies. We show how their abundance ratios compare with low- and high-redshift  galaxies, discuss how these galaxies behave in the BPT diagram \citep{Baldwin1981Classification-}, and how the N-abundance affects this diagram.

The structure of our paper is as follows. In Sect.~\ref{s_obs} we briefly describe our JWST observations, data reduction, and measurements, as well as comparison samples.  The methods adopted to determine elemental abundances are described in Sect.~\ref{s_abund}.
The BPT diagram of the LyC22 sample is shown and discussed in Sect.~\ref{s_bpt}. In  Sect.~\ref{s_results} we present the derived abundances and discuss the behaviour of N/O. Our main conclusions are summarized in Sect.~\ref{s_conclude}.


\begin{figure*}[htb]
\centering
\includegraphics[width=1\textwidth]{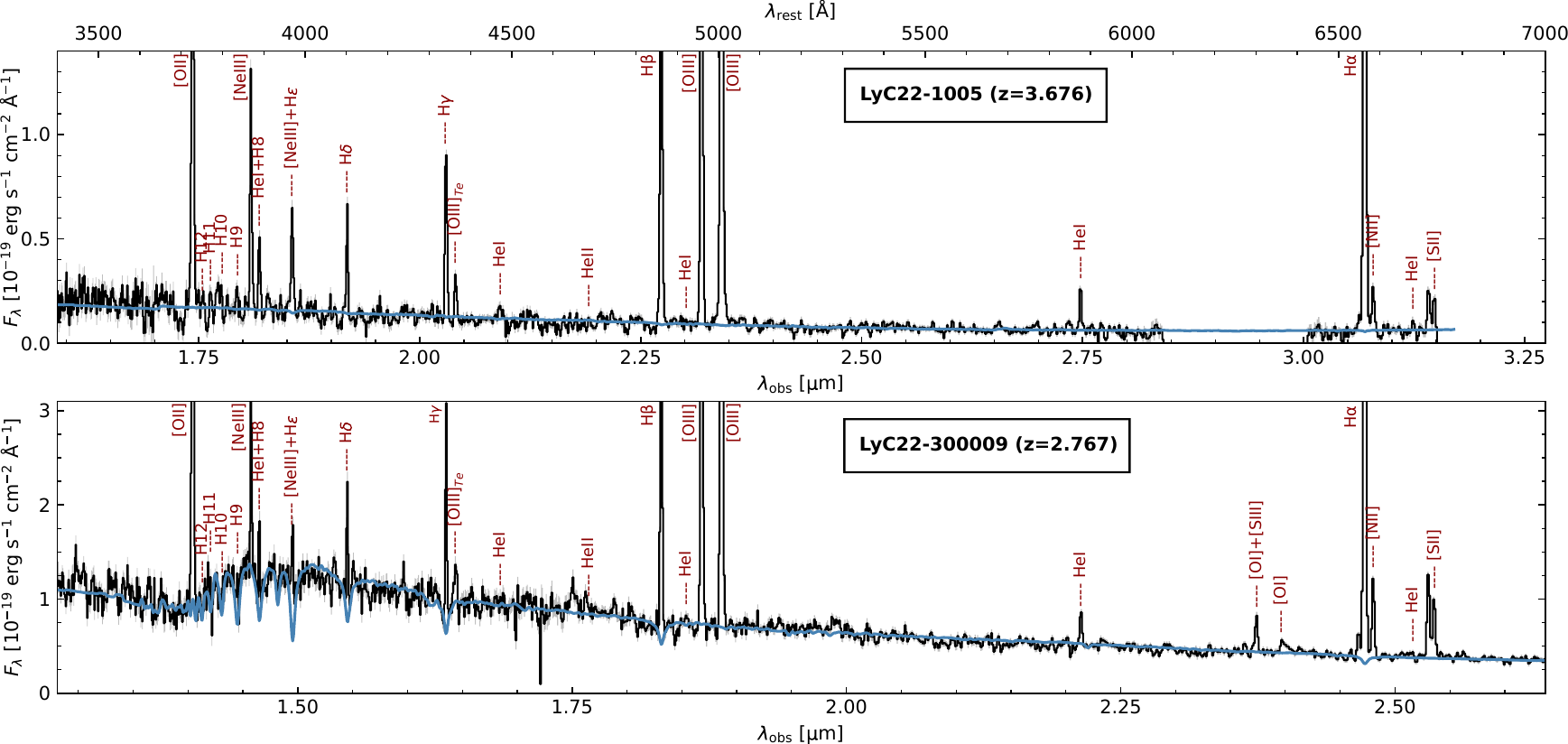}
\caption{JWST spectra from LyC22 showing object 1005 (top panel, LACES $z=3.676$ LAE in SSA22) and 300009 (bottom, $z=2.767$ LBG in the Westphal field). The blue line shows the adopted fit to the continuum. The main emission and absorption lines are indicated.}
\label{fig_spec}
\end{figure*}


\section{Observations}
\label{s_obs}

The LyC22 survey 
is a JWST spectroscopic program to test indirect indicators of Lyman continuum escape, study the interstellar medium, radiation field, and other properties of known Lyman continuum emitters and a control sample of $z \sim 3$ galaxies, building on the KLCS and LACES surveys \citep{Steidel2018The-Keck-Lyman-,Fletcher2019The-Lyman-Conti,Pahl2021An-uncontaminat}. 
These surveys have mostly included  narrow-band selected \lya\ emitters and Lyman break galaxies at $z \sim 2.5-3.5$, which represent the main populations of star-forming galaxies at these redshifts \citep[e.g.][]{Shapley2011Physical-Proper}. We therefore refer to these galaxies as  'typical' star-forming galaxies at this epoch.

Targeting a mixture of both narrow-band selected \lya\ emitters and Lyman break galaxies at $z \sim 2.5-3.5$  in the SSA22 and  Westphal fields observed by the KLCS and LACES surveys,  the final LyC22 observations have provided 117 spectra from three pointings,
with a median spectroscopic redshift $z=2.95^{+0.32}_{-0.73}$
 (for all quantities we here quote the median and 16, 84\% percentiles of the sample.)
The LyC22 targets have stellar masses $\mstar \sim 10^{8.3}$ to $10^{10.7}$ \msun,  and star-formation rates SFR $\sim 1-100$ \msunyr, as derived from standard SED fits assuming constant SFR. More detailed properties of the sample will be described in a subsequent publication.

\subsection{JWST NIRSpec observations}
The LyC22 sources were observed with NIRSpec in three different micro-shutter assembly (MSA) pointings (two in the SSA22, one in Westphal). 
Each MSA slitlet consisted of three micro-shutters.
We used the medium-resolution grating/filter combinations G140M/F100LP and G235M/F170LP, delivering a spectral resolving power of $R \sim 1000$ and continuous wavelength coverage from 0.97 to 3.07\,$\mu$m. Each source received total on-source integration times of $\simeq 9.2$\,hours (G140M) and $\simeq 8.8$\,hours (G235M), distributed over 18 individual integrations using the NRSIRS2 readout mode. We reached a $3\sigma$ median line flux sensitivity of $\simeq 2.5\times10^{-19}$ and $\simeq 1.8\times10^{-19}$ erg\,s$^{-1}$\,cm$^{-2}$ in the blue and red gratings (for a line width of 300\,km\,s$^{-1}$). 

Data reduction was performed using the official \textit{JWST} calibration pipeline (v1.13.4) for Level~1 data, and the \texttt{msaexp}\footnote{\url{https://github.com/gbrammer/msaexp}} package (v0.7.3; \citealt{msaexp}) for Levels~2 and~3. Processing steps included subtraction of bias and dark current, correction for 1/$f$ noise, and detection and masking of cosmic ray snowballs. We adopted the calibration reference data system (CRDS) context \texttt{jwst\_1202.pmap} to apply flat-field corrections and perform wavelength and flux calibration. The 2D spectra from each slitlet were drizzle-combined, and background subtraction was applied using the standard three-shutter nod pattern. 1D science and error spectra were extracted using the optimal extraction method of \cite{horne1986} from the rectified G235M 2D spectra, and the same extraction parameters were applied to the G140M data. 
The error spectra were rescaled upwards to match the observed rms. The 1D spectra were resampled onto a common wavelength grid and flux-matched in the overlapping region ($\lambda \simeq 1.73$--$1.85\,\mu$m).
Finally, we followed the approach described in \citet{Reddy2023} to determine wavelength-dependent slit losses, using a fit of the light profile from the NIRCam F277W images and comparing the fraction of light captured within the slit and extraction apertures as a function of wavelength.

The deep medium-resolution NIRSpec observations provide a full spectral coverage over $\sim 2800-6800$ \AA\ in the rest frame, including the major indicators of LyC escape recently established from studies at low-$z$ \citep[cf.][]{Flury2022bThe-Low-redshif,Jaskot2024Multivariate-I} and numerous other emission lines of interest. 
Spectra of two LyC22 star-forming galaxies with clear \Oiiit\ auroral line detections (discussed below) are shown in Fig.~\ref{fig_spec} for illustration.
%

\subsection{Emission line measurements}

We modeled the stellar continuum of each LyC22 source using the \texttt{pPXF} software \citep{Cappellari2017}, allowing us to account for underlying stellar absorption, particularly beneath Balmer lines (see Fig.~1). The fits were performed using \texttt{fsps} stellar population synthesis models \citep{Conroy2010FSPS:-Flexible-}, allowing for ages between 1\,Myr and 2.2\,Gyr.
After continuum subtraction, emission line fluxes were measured using \texttt{LiMe} \citep{LiMe}, fitting Gaussian profiles within a rest-frame spectral window of 20\,\AA\ centered on the expected wavelength to the spectra, and using tailored continuum windows.
Closely spaced emission lines, such as H$\alpha$ and [N~{\sc ii}] $\lambda \lambda 6550,6585$, were fitted simultaneously with multiple Gaussians. We measured a comprehensive set of emission lines, from a list of more than 40 features from [Ne~{\sc iv}] $\lambda2423$ to [Ar~{\sc iii}] $\lambda7751$. 
Finally, emission line fluxes were corrected for extinction using the Balmer decrement. 
We used the main Balmer lines (H$\alpha$, H$\beta$, H$\gamma$, and H$\delta$) and adopted the extinction curve from \citet{Cardelli1989The-relationshi}, assuming standard nebular conditions ($n_e = 100$\,cm$^{-3}$ and $T_e = 10^{4}$).
The median extinction of the LyC22 sources is low, $E(B-V) = 0.008^{+0.24}_{-0.007}$.

\subsection{Comparison samples}

We also compare the LyC22 sources with galaxies from the low-$z$ Lyman Continuum Survey (LzLCS) at $z \sim 0.3$ \citep{Flury2022aThe-Low-redshif,Flury2022bThe-Low-redshif}. This sample includes 66 star-forming galaxies observed with Hubble in the UV and Lyman continuum, selected to span a range in UV slopes, SFR surface densities, and \Oiii/\Oii\ ratios \citep{Flury2022aThe-Low-redshif}, and 23 compact star-forming galaxies observed earlier by Izotov and collaborators \citep[e.g.][]{Izotov2016Eight-per-cent-,Izotov2018Low-redshift-Ly}.
The resulting sample includes, for example, galaxies with stellar masses $\mstar \sim 10^{7.3}-10^{10.5}$ \msun, SFR(UV) $\sim 1-100$ \msunyr, and SFR surface densities $\log(\Sigma_{\rm SFR}) \sim -1$ to 2 \msunyr kpc$^{-2}$.  Although relatively heterogenous, the low-$z$ comparison sample (89 SFGs) is overall biased towards low metallicity and more actively star-forming galaxies, and does not represent the full distribution of $z \sim 0$ star-forming galaxies from the SDSS. 

In terms of stellar mass and SFR, the LzLCS and LyC22 show a strong overlap (cf.~above), although the low-$z$ sample also includes 10 objects with masses $\mstar \sim 10^7-10^8$ \msun, below the range covered by LyC22. Further overlaps between these samples will be shown below.

Finally, we also compare our results with other JWST observations from the literature, in particular with other studies discussing the nitrogen abundance and N/O in star-forming galaxies, including results from programs such as MARTA, AURORA, EXCELS, and others \citep[][]{Arellano-Cordova2024The-JWST-EXCELS,Sanders2025The-AURORA-Surv,Scholte2025The-JWST-EXCELS,Stiavelli2025What-Can-We-Lea,Cataldi2025Tracing-Nitroge}.
The properties of the targets of these programs are very diverse, and the selection criteria generally not very well constrained.

\begin{figure}[tb]
\includegraphics[width=0.5\textwidth]{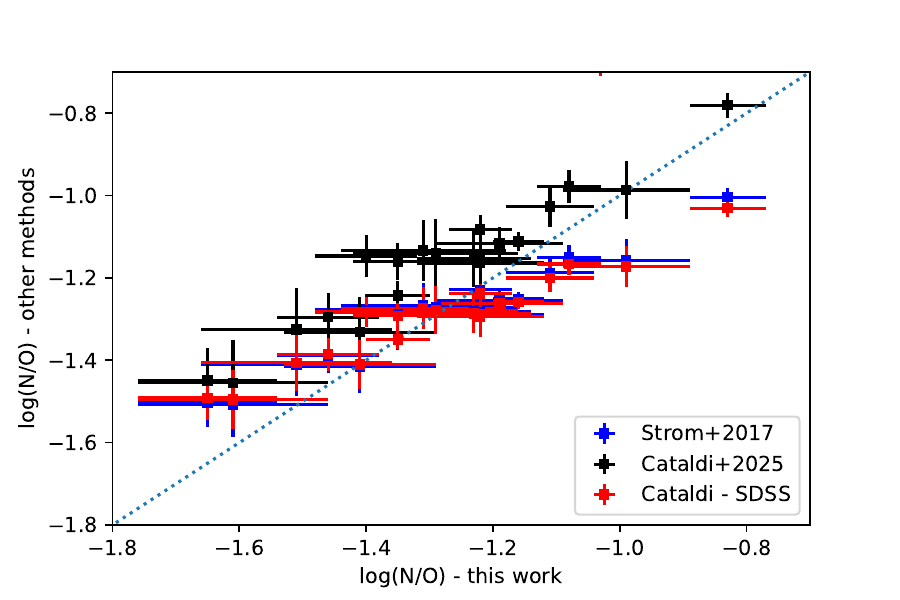}
\caption{Comparison of the N/O abundances derived following the direct method and the assumptions described in this work (x-axis) with those adopting the  strong line calibrations of \cite{Cataldi2025Tracing-Nitroge} based on JWST samples (in black; applicable for $z>1$) and an SDSS sample (red, applicable at low-$z$), and the calibration from \cite{Strom2017Nebular-Emissio} (blue). All strong line calibrations use the extinction-corrected line ratio of \Nii/\Oii.}
\label{fig_compare_no}
\end{figure}

\begin{figure*}[tb]
\includegraphics[width=0.5\textwidth]{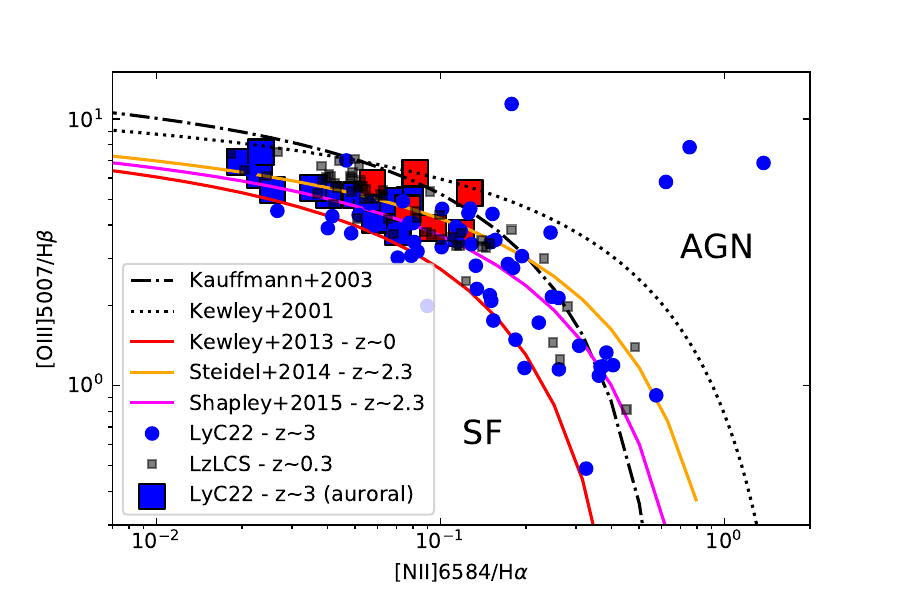}
\includegraphics[width=0.5\textwidth]{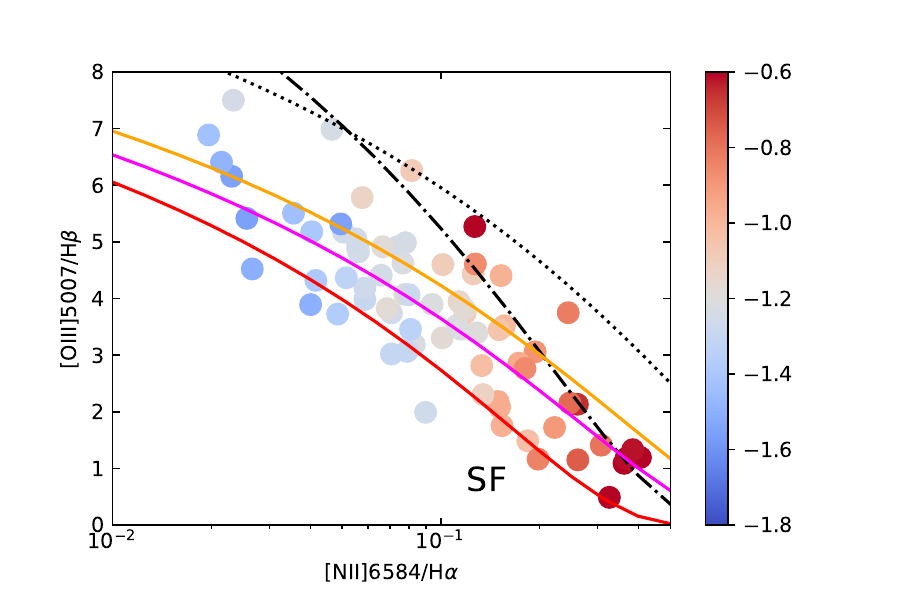}
\caption{ {\em Left: } Classical emission line diagnostic diagram showing the galaxies from the LyC22 sample (blue and red symbols). 
LyC22 objects with significant \Oiiit\ detections are surrounded by a black square. LyC22 galaxies with $\log({\rm N/O})>-1.2$ are shown in red.
We only show objects where the involved emission lines are detected at  $\ge 3 \sigma$. Typical uncertainties are comparable to the size of the symbols, hence not plotted.
The maximum starburst line from \cite{Kewley2001Optical-Classif} (dotted) and the empirical AGN/star-formation threshold from \cite{Kauffmann03} established for $z \sim 0$ (dash-dotted) are also shown. Average relations for $z \sim 2.3$ galaxies from \citep{Steidel2014Strong-Nebular-,Shapley2015The-MOSDEF-Surv} are shown by orange and magenta lines, average relation of SDSS galaxies from \cite{Kewley2013The-Cosmic-BPT-} by the red line.
Observations from the LzLCS at $z\sim 0.3$ are shown by small grey squares.
{\em Right:} Zoom on part of the BPT diagram showing all LyC22 sources with color-coded N/O abundances  (in $\log($N/O$)$), computed from \nii/\oii.
}
\label{fig_bpt}
\end{figure*}

\section{Nebular abundance determination}
\label{s_abund}

Since accurate abundance determinations require auroral lines, we examine all objects with potential \Oiiit\ detections. After eliminating the AGN identified previously and careful visual inspection, this leads to a final sample of 25 galaxies with robust \Oiiit\ detections.  We then follow  \cite{Izotov2006The-chemical-co} to determine abundances, using the direct method.

Electron densities, measured from the \Sii\ doublet, are relatively low ($n_e= 152^{+425}_{-142}$ \cmc ), and  electron temperatures not unusual ($T_{\rm e}$(O~{\sc iii})$=(1.42^{+0.21}_{-0.24})\times 10^4$ K) for subsolar metallicity star-forming regions. They also agree with those from other studies at $z \sim 1-3$ using other JWST observations \citep[see e.g.][]{Sanders2025The-AURORA-Surv,Cataldi2025MARTA:-Temperat}.

The electron temperature $T_{\rm e}$(O~{\sc iii}) is used to obtain the ionic abundances of O$^{2+}$ and Ne$^{2+}$; the temperature in the low-ionization region, $T_{\rm e}$(O~{\sc ii}) obtained from the former as described by \cite{Izotov2006The-chemical-co},  was used to derive the ionic abundance of O$^+$ and N$^{+}$.
For the total oxygen abundance O/H we use the ionic abundances of  O$^{2+}$/H$^+$ and O$^{+}$/H$^+$, finding that the former generally dominates in our objects. For N/H we used the optical \Nii\  lines and the ionization correction factor (ICF) from \citet{Izotov2006The-chemical-co}; similarly for Ne/H using \Neiii.

For the LzLCS sample we have derived N/O from the  extinction-corrected \protect\nii/\oii\ ratio,  following \cite{Strom2017Nebular-Emissio}, and taken their metallicities (O/H) from the original papers \citep{Flury2022aThe-Low-redshif,Flury2022bThe-Low-redshif}.  
For the majority of  LzLCS  objects, O/H was  derived using the direct method.

Finally, we have also compared our abundance determinations to those using the direct method but with other prescriptions, e.g., for the $T_{\rm e}($O~{\sc ii}$)$ which is not directly measured, or for the ICF of N/O. 
In Fig.~\ref{fig_compare_no} we compare the N/O abundances derived for the LyC22 using our method to those obtained from the extinction-corrected \Nii/\Oii\ line ratio with the  prescription from \cite{Strom2017Nebular-Emissio} and the more recently derived fits from \cite{Cataldi2025Tracing-Nitroge}, who provide strong-line calibrations for N/O using a high-$z$ galaxy sample observed with JWST and an SDSS sample. 
As Fig.~\ref{fig_compare_no} shows, our N/O measurements correlate well with their JWST-based calibration, with an offset of $-0.105$ dex on average.
Furthermore, we also see that their SDSS calibration agrees very well with that of  \cite{Strom2017Nebular-Emissio}, which we use for the LzLCS comparison sample, confirming thus further the use of this relation for our low-$z$ sources.
Further discussions and comparisons of methods to determine N/O abundances from the optical line ratios are discussed in \cite{Cataldi2025Tracing-Nitroge}.
In short, we conclude that our standard methodology is in good agreement with other studies, with the caveat of a systematic shift in the N/O abundance of $\sim -0.1$ dex compared to the N/O strong line calibration of \cite{Cataldi2025Tracing-Nitroge}, which does, however, not affect our conclusions (see below).

\section{BPT diagram of galaxies at $z \sim 3$}
\label{s_bpt}

In Fig.~\ref{fig_bpt} we show the position of the LyC22 sources in one of the classical emission-line (so-call BPT) diagnostic diagrams from \cite{Baldwin1981Classification-}, together with the ``maximum starburst'' line from \cite{Kewley2001Optical-Classif} and an empirical AGN/star-formation threshold from \cite{Kauffmann03} established for $z \sim 0$ galaxies from the SDSS.
Based on this diagram we robustly identify four LyC22 sources as AGN, which is also confirmed by their broad hydrogen emission lines. The remainder of our objects are found below the discrimination line of \cite{Kewley2001Optical-Classif}, and the majority also below the line from \cite{Kauffmann03}, therefore compatible with photoionization by massive stars. 
Interestingly, these SF-AGN separations established for $z \sim 0$ galaxies also suffice to describe the envelope of our $z \sim 3$ galaxies, and do not require shifted separation lines for $z \sim 3$, as proposed earlier by several studies \cite[e.g.][]{Kewley2013Theoretical-Evo,Kewley2013The-Cosmic-BPT-}.

The LyC22 SFGs are in good agreement with the average relations established earlier for $z \sim 2.3$ galaxies with Keck observations  \citep{Steidel2014Strong-Nebular-,Shapley2015The-MOSDEF-Surv}. This confirms, for the first time with a large sample at $z \sim 3$, the offset of high-$z$ SFGs found previously for $z\sim 2.3$ from the average relation of SDSS galaxies, as illustrated by the $z \sim 0$ line from \cite{Kewley2013The-Cosmic-BPT-}.
Benefiting from accurate abundances for a subset of our galaxies, we also find that galaxies with higher-than-average N/O ratios have systematically higher \nii/\ha\ ratios at a given \oiii/\hb, as shown by the red squares in Fig.~\ref{fig_bpt}  (left panel). Indeed, as shown in the right panel,  galaxies with  increasing N/O abundance are shifted to higher \nii/\ha\ ratios at a given \oiii/\hb, as expected from simple nebular physics, and also observed  in SDSS samples \citep{Masters2016A-Tight-Relatio}.

\begin{figure}[htb]
\includegraphics[width=0.5\textwidth]{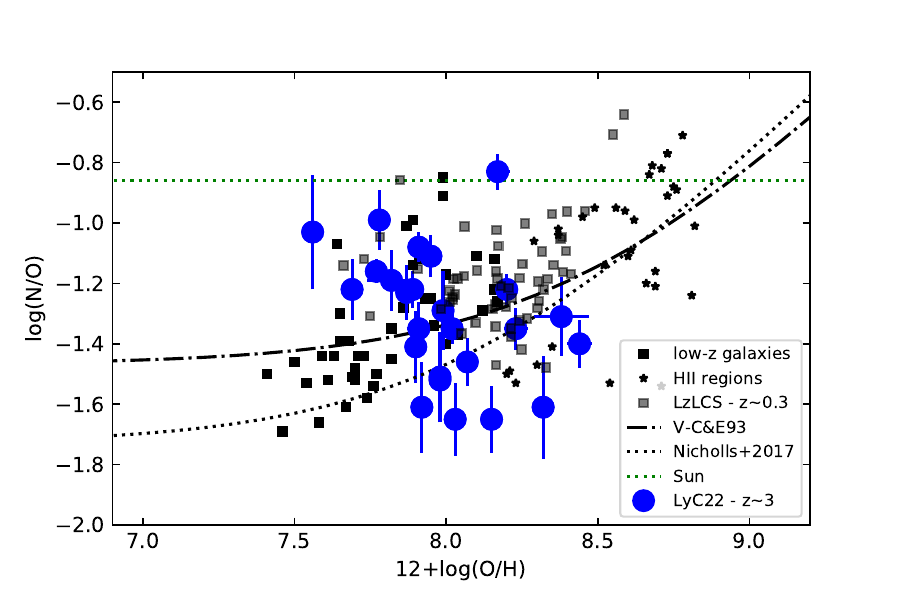}
\caption{Derived abundance ratio N/O from rest-optical lines of the LyC22 galaxies (blue circles) and low-$z$ samples as a function of O/H. The low-$z$ star-forming galaxies and \hii\ regions from the compilation of \cite{Izotov2023Abundances-of-C} and the LzLCS  are shown by small black and grey symbols, respectively. Dash-dotted and dotted lines show the average trend observed in low-$z$ star-forming galaxies, as parametrized by \cite{Vila-Costas1993The-nitrogen-to}  and \cite{Nicholls2017Abundance-scali} respectively. The green dotted line shows the solar value. }
\label{fig_abund}
\end{figure}

We also compare the LyC22 sources with the low-$z$ Lyman Continuum Survey (LzLCS) at $z \sim 0.3$ in Fig.~\ref{fig_bpt}. 
They follow very similar sequences in this BPT diagram, showing thus  similar excitation properties for the LyC22 and LzLCS samples.
Furthermore, since the inferred N/O abundances (cf.~below) are also comparable between these $z \sim 0$ and $z \sim 3$ samples and not significantly enhanced, this implies that other causes explain the offset in the BPT diagram between the overall $z \sim 0$ population and $z \sim 2-3$ SFGs.
For example, \cite{Steidel2016Reconciling-the} and \cite{Strom2017Nebular-Emissio} concluded that a harder ionizing spectrum at fixed O/H, due to iron-poor stellar populations,  is the most likely explanation for the offset.
Alternatively,  \cite{Masters2016A-Tight-Relatio} have shown that low-$z$ galaxies with higher SFR surface densities are shifted towards the locus of $z \sim 2-3$ galaxies, which is also reflected in the shift between the overall (average) SDSS population and the LzLCS $z \sim 0.3$, by selection (see above).
The two explanations could in fact be consistent, since alpha-enhancements (increased O/Fe) are equivalent to Fe-deficiencies at fixed O/H, and found to increase with increasing  specific SFR (sSFR), which also correlates with $\Sigma_{\rm SFR}$ \citep[e.g.][]{Masters2016A-Tight-Relatio,Flury2022bThe-Low-redshif}.

\section{Normal nitrogen and oxygen abundances in star-forming galaxies at $z \sim 3$}
\label{s_results}

The  N/O ratios and metallicities (O/H) of our sample are shown in Fig.~\ref{fig_abund}.
O/H ranges from $\oh=7.56$ to 8.44 (with a median of 7.93), the lowest metallicity thus being $\sim 7$\% solar \citep{Asplund2009The-Chemical-Co}.
With $\log({\rm N/O})=-1.29^{+0.25}_{-0.22}$ the observed N/O ratios of the LyC22 sample are thus clearly subsolar, and both the average N/O ratio and the observed scatter broadly agree with those found in low-$z$ star-forming galaxies and \hii\ regions at the same metallicity, as also shown in this figure.
For metallicities $\oh<8.45$ comparable to LyC22, the  \cite{Izotov2023Abundances-of-C} sample shown here has $\log({\rm N/O})=-1.33 \pm 0.21$, and the LzLCS  $\log({\rm N/O})=-1.22 \pm 0.18$.

We do not consider that the apparent trend of decreasing N/O with increasing O/H is significant\footnote{Both Spearmann and Kendall's test yield comparable values of $p=0.02$.}. 
Comparable galaxies are also found at low-$z$ and the number of objects at $\oh \la 7.8$ is quite limited.  
Our sample also does not show a correlation of N/O with density, which, according to \cite{Arellano-Cordova2025CLASSY-XIV:-Nit}, could explain the observed scatter of N/O.
The  scatter in the N/O-O/H relation is well known and can in principle be explained in terms of different timescales over which nitrogen and oxygen are released into the ISM \citep[see e.g.][]{Perez-Montero2009The-impact-of-t,Berg2019The-Chemical-Ev}, different star formation histories \citep[e.g.][]{Henry2000On-the-Cosmic-O}, or other processes, although no dominant mechanism has been identified \citep[see e.g. discussions in ][]{Arellano-Cordova2025CLASSY-XIV:-Nit,Cataldi2025MARTA:-Temperat}.
However, it has been shown that at a given metallicity, N/O does not correlate with stellar mass \citep[e.g.][]{Masters2016A-Tight-Relatio,Arellano-Cordova2025CLASSY-XIV:-Nit}.

In any case, our main empirical finding is that typical star-forming galaxies at $z \sim 3$ have essentially the same N/O ratio (and scatter) at a given O/H, as galaxies and \hii\ regions at low-redshift. 
With 25 new measurements, our sample increase the number of known galaxies at $z \sim 2-4$ with accurate N/O measurements from JWST using the same method as here (direct method and optical \nii\ lines). This result is also corroborated by the recent study of \cite{Cataldi2025MARTA:-Temperat}, who add other data.

We also examined the neon abundance ratio, finding $\log({\rm Ne/O}) = -0.73^{+0.11}_{-0.08}$, also in good agreement with 
 low-redshift star-forming galaxies \citep[e.g.][]{Guseva2011VLT-spectroscop}. 

\begin{figure}[tb]
\includegraphics[width=0.5\textwidth]{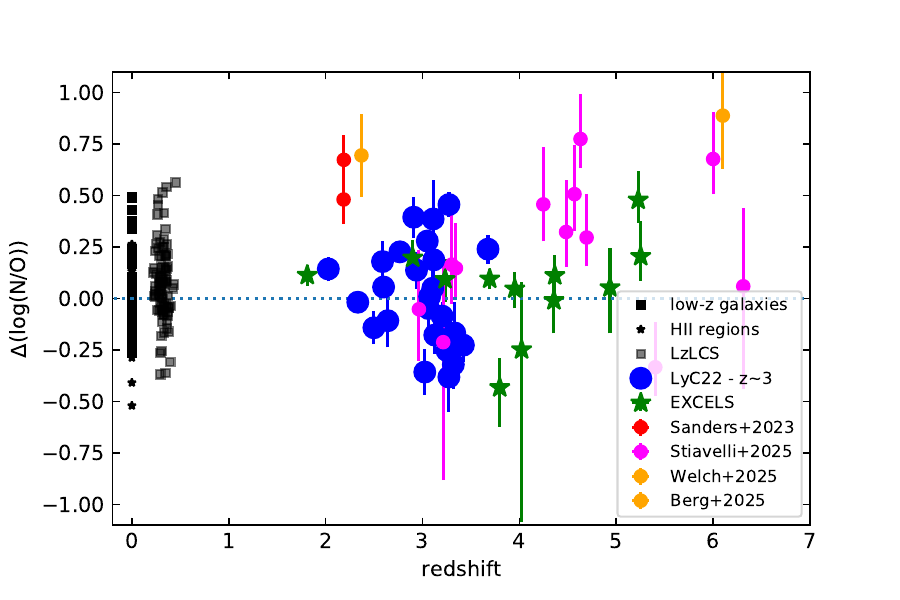}
\caption{Excess  $\Delta(\log({\rm N/O}))$ of N/O with respect to the local N/O--O/H relation from \cite{Vila-Costas1993The-nitrogen-to} as a function of redshift for galaxies with abundances from the direct method and N/O from rest-optical lines.
Low-$z$ star-forming galaxies and \hii\ regions, LzLCS and LyC22 samples are shown with the same symbols as in Fig.~\protect\ref{fig_abund}.
Other N/O measurements from the rest-optical lines, obtained primarily from JWST spectra, are shown at $z \sim 2.2$ \citep[red symbols, from][]{Sanders2023A-Preview-of-JW}, for the EXCELS sample \citep[green stars;][]{Arellano-Cordova2024The-JWST-EXCELS,Scholte2025The-JWST-EXCELS}, from \cite[][magenta]{Stiavelli2025What-Can-We-Lea}, the Sunburst arc at $z=2.4$ \citep[orange;][]{Welch2025The-Sunburst-Ar}, and for the $z=6.1$ lensed galaxy (orange) from \cite{Topping2024Metal-poor-star}, with measurements from rest-optical lines by \cite{Berg2025A-Fleeting-GLIM}. The two orange symbols indicate objects known to show also UV emission lines of Nitrogen.}
\label{fig_no_z}
\end{figure}

\subsection{Does the average N/O ratio at low metallicity increase beyond $z \protect\ga 4$ ?}

To examine if at higher redshift ($z \ga 4$) the average N/O ratio increases and deviates from the average relation observed at $z \sim 0-3$, as suggested e.g.~by  \cite{Hayes2025On-the-Average-} and \cite{Cataldi2025Tracing-Nitroge}, we show in Fig.~\ref{fig_no_z} the difference 
$\Delta(\log({\rm N/O})) = \log({\rm N/O}) - \log({\rm N/O})_{\rm VCE}$  between the derived N/O ratio and the value expected from the relation of \cite{Vila-Costas1993The-nitrogen-to} at the observed O/H.  For comparison we show samples with available abundances determined from JWST spectra using the direct ($T_e$) method and N/O ratios derived from the rest-optical lines, i.e.~using the same methodology as our study. This also includes three objects at $z \sim 2.2-2.4$ studied by \cite{Sanders2023A-Preview-of-JW} and \cite{Welch2025The-Sunburst-Ar}, and the lensed N-emitter at $z=6.1$ identified by \cite{Topping2024Metal-poor-star}.

As shown in Fig.~\ref{fig_no_z}, N/O shows a large dispersion, also at $z \ga 4$, and the average N/O ratio is somewhat enhanced, with respect to the ``local'' N/O--O/H relation, although the current samples with accurate abundances are still small. Qualitatively the same is also found if we used, e.g., the more recent local N/O--O/H relation determined by \cite{Cataldi2025Tracing-Nitroge}. 
Clearly, a few individual objects with significant N/O excess measured from the optical lines exist, both at $z \sim 4-6$ and also at $z \sim 2.2$, such as the well-studied Sunburst arc \citep{Pascale2023Nitrogen-enrich,Welch2025The-Sunburst-Ar} and others.
How high and how significant the excess of the average N/O ratio is, and how this evolves at $z \ga 4$ requires accurate abundance determinations in larger samples of  typical galaxies at high-$z$, which remains largely to be done.

In this context it is also interesting to compare our findings with those from studies of rare galaxies showing nitrogen lines (\Nivuv\ or \Niiiuv) in the UV domain, which we refer to as N-emitters. 
Fig.~\ref{fig_no_z} shows two such objects, the Sunburst arc at $z=2.4$ and RXCJ2248-ID3 at $z=6.1$ shown in orange (with an excess $\Delta(\log({\rm N/O})) \sim 0.7-0.9$), which both also have a high N/O ratio according to the optical line ratios. To the best of our knowledge no other of the $z >2$ galaxies plotted in Fig.~\ref{fig_no_z} show reported N-lines in the UV. This is in part due to the absence of rest-UV spectra of these objects, and to the non-detection of these lines. For the $z \sim 2-3$ galaxies including the LyC22 targets, in particular, deep ground-based rest-UV spectra have also not shown any detections of the \Nivuv\ or \Niiiuv\ lines, in agreement with earlier studies, which have shown that these lines are generally not present in UV spectra of $z \sim 2-3$ star-forming galaxies, and only very rarely found individually or in stacks of extreme objects \citep[e.g.][]{Christensen2012The-low-mass-en,Le-Fevre2019The-VIMOS-Ultra}.
Our finding of a ``normal'' N/O ratio in ``typical'' $z \sim 3$ galaxies is fully consistent with this picture, where only very few objects show a strong excess in N/O at these redshifts. According to the recent work of \cite{Morel2025Discovery-of-ne} and \cite{Schaerer2025N-emitters-as-p}, the fraction of star-forming galaxies showing UV nitrogen lines is $f_N \sim (1-3) \times 10^{-3}$ at  $z \sim 3$, too low to show up and affect samples like those studied here.
Whether the observed increase of $f_N$ with redshift is compatible with and sufficient to explain the possible increase of the average N/O ratio at $z \ga 4$ discussed here, remains to be examined.

\section{Conclusions}
\label{s_conclude}

We have presented the first JWST spectra from the LyC22 survey, observing known Lyman-continuum emitters and a control sample at $z \sim 3$, selected primarily from earlier LyC studies with HST and Keck \citep{Steidel2018The-Keck-Lyman-,Fletcher2019The-Lyman-Conti,Pahl2021An-uncontaminat}, which targeted a representative sample of LAEs and LBGs.

Using LyC22 spectra with robust detections of the \Oiiit\ auroral line, we have determined accurate chemical abundances of N, O, and Ne relative to H for a subset of 25 star-forming galaxies from their rest-optical emission lines.
Their metallicities range between $\oh \sim 7.5$ and $\sim 8.4$ and they all show sub-solar N/O abundance ratios $\log({\rm N/O})=-1.29^{+0.25}_{-0.22}$, which is in agreement with the observed N/O abundances of low-$z$ galaxies and \hii\ regions at the same metallicity. 
From our data and other existing JWST measurements we therefore conclude that typical star-forming galaxies (LAEs, LBGs) at $z \sim 3$ are not enriched in nitrogen, compared to their low-redshift counterparts  at  given metallicity O/H.
Statistics for larger samples and establishing the distribution of N, O, and other elemental abundances, in particular at $z \ga  4$, will be of great interest for our understanding of the early chemical evolution of galaxies and  the nature of  rare N-emitters discovered by JWST.

Our spectra show that the LyC22 sample closely resembles those of other $z \sim 2$ samples in the \nii/\ha\ BPT diagram, confirming earlier findings of an offset between SFGs at these redshifts and the average of SDSS galaxies \citep{Steidel2014Strong-Nebular-,Shapley2015The-MOSDEF-Surv}. 
From our abundance determinations we found that galaxies with higher N/O ratios show stronger \nii/\ha\ emission, as expected. However, our results imply that observed shifts in the BPT are not due to enhanced N-abundances, as suggested by some earlier studies \citep[e.g.][]{Shapley2015The-MOSDEF-Surv,Masters2016A-Tight-Relatio}.
Populations with similar offsets also exist at low-$z$ (e.g.~the LzLCS sample), and their offset could be due to higher SFR surface densities or a harder ionizing spectrum at fixed O/H \citep[cf.][]{Masters2016A-Tight-Relatio,Steidel2016Reconciling-the}.

\begin{acknowledgements}
Y.I., N.G, R.M.-C., and D.S. acknowledge support from project No. 224866 carried out in the framework of ``Ukrainian-Swiss Joint Research Projects: Call for Proposals 2023''.

\end{acknowledgements}
\bibliographystyle{aa}



\end{document}